\documentclass[conference]{IEEEtran}
\IEEEoverridecommandlockouts

\usepackage[margin=1in]{geometry}
\usepackage[utf8]{inputenc}
\usepackage{graphicx}
\usepackage{hyperref}
\usepackage{amsmath}
\usepackage{amssymb}
\usepackage[acronym]{glossaries}
\usepackage{multirow}
\usepackage{stackengine,scalerel}
\usepackage{accents}
\usepackage{algorithm}
\usepackage{algorithmic}
\usepackage{gensymb}
\usepackage{cite}
\usepackage{amsmath,amssymb,amsfonts}
\usepackage{graphicx}
\usepackage{textcomp}
\usepackage{xcolor}
\usepackage{optidef}
\usepackage{multirow}
\usepackage{stfloats}
\usepackage{lipsum}
\usepackage{subcaption}
\usepackage{optidef}
\usepackage{cite}
\usepackage{amsmath,amssymb,amsfonts}
\usepackage{algorithmic}
\usepackage{graphicx}
\usepackage{textcomp}
\usepackage{xcolor}
\usepackage{stfloats}  
\usepackage{cuted}     
\usepackage{subcaption}
\usepackage{color}

\def\BibTeX{{\rm B\kern-.05em{\sc i\kern-.025em b}\kern-.08em
    T\kern-.1667em\lower.7ex\hbox{E}\kern-.125emX}}

\renewcommand{\a}{\mathbf{a}}
\renewcommand{\b}{\mathbf{b}}

\newcommand{\e}{\mathbf{e}}
\newcommand{\f}{\mathbf{f}}

\newcommand{\h}{\mathbf{h}}
\renewcommand{\i}{\mathbf{i}}

\newcommand{\q}{\mathbf{q}}

\newcommand{\A}{\mathbf{A}}
\newcommand{\B}{\mathbf{B}}
\newcommand{\C}{\mathbf{C}}

\newcommand{\F}{\mathbf{F}}

\renewcommand{\H}{\mathbf{H}}

\newcommand{\U}{\mathbf{U}}

\newcommand{\W}{\mathbf{W}}

\newcommand{\Piv}{\boldsymbol{\Pi}}

\newcommand{\setK}{\mathcal{K}}

\newcommand{\herm}{\mathrm{H}}
\newcommand{\tr}{\mathrm{Tr}}
\newcommand{\tran}{\mathrm{T}}

\renewcommand{\rm}[1]{\mathrm{#1}}
\newcommand{\norm}[1]{\left\lvert\left\lvert#1\right\rvert\right\rvert}
\newcommand{\abs}[1]{\left\lvert#1\right\rvert}

\newcommand{\qed}{\nobreak \ifvmode \relax \else
      \ifdim\lastskip<1.5em \hskip-\lastskip
      \hskip1.5em plus0em minus0.5em \fi \nobreak
      \vrule height0.75em width0.5em depth0.25em\fi}

\begin{document}

\glsdisablehyper   

\newacronym[]{AoD}{AoD}{angle of departure}
\newacronym[]{AoA}{AoA}{angle of arrival}
\newacronym{ADC}{ADC}{Analog-to-Digital Converter}
\newacronym{AO}{AO}{alternating optimization}
\newacronym{BB}{BB}{baseband}
\newacronym[]{BF}{BF}{beamforming}
\newacronym{BS}{BS}{base station}
\newacronym[]{CRB}{CRB}{Cramer-Rao bound}
\newacronym{CRSM}{CRSM}{cluster ray sensing model}
\newacronym{DoF}{DoF}{Degree of Freedom}
\newacronym{DL}{DL}{downlink}
\newacronym{DoA}{DoA}{direction of arrival}
\newacronym{DoD}{DoD}{direction of departure}
\newacronym{DFT}{DFT}{discrete fourier transform}
\newacronym{DFRC}{DFRC}{dual-function radar-communication}
\newacronym{FD-ISAC}{FD-ISAC}{full duplex integrated sensing and communication}
\newacronym{FOV}{FOV}{field of view}
\newacronym{5G}{5G}{fifth generation}
\newacronym{FD}{FD}{full duplex}
\newacronym{FDD}{FDD}{frequency division duplex}
\newacronym{FMCW}{FMCW}{frequency-modulated continuous-wave}
\newacronym{HBF}{HBF}{Hybrid beamforming} 
\newacronym{HD}{HD}{half duplex}
\newacronym[]{ISAC}{ISAC}{Integrated Sensing and Communication}
\newacronym{IID}{IID}{independent and identically distributed}
\newacronym{LOS}{LOS}{line of sight}
\newacronym{MUSIC}{MUSIC}{MUltiple SIgnal Classification}
\newacronym[]{MPC}{MPC}{multi path component}
\newacronym[]{MIMO}{MIMO}{multiple input multiple output}
\newacronym[]{mmWave}{mmWave}{millimeter-Wave}
\newacronym{MSS}{MSS}{maximum signal strength}
\newacronym{MSE}{MSE}{mean square error}
\newacronym{NLOS}{NLOS}{non line of sight}
\newacronym{NSP}{NSP}{null space projection}
\newacronym{NR}{NR}{new radio}
\newacronym{OFDM}{OFDM}{orthogonal frequency division multiplexing}
\newacronym{PAPR}{PAPR}{Peak-to-Average-Power-Ratio}
\newacronym{PSD}{PSD}{positive semi-definite}
\newacronym{QoS}{QoS}{Quality-of-service}
\newacronym{RSU}{RSU}{road side unit}
\newacronym[]{RX}{RX}{receiver}
\newacronym[]{SINR}{SINR}{signal-to-interference plus noise ratio}
\newacronym{SI}{SI}{self-interference}
\newacronym{SNR}{SNR}{signal to noise ratio}
\newacronym{SDR}{SDR}{Semi-definite relaxation}
\newacronym{SRSM}{SRSM}{single ray sensing model}
\newacronym[]{TX}{TX}{transmitter}
\newacronym{ULA}{ULA}{uniform linear array}
\newacronym{USA}{USA}{Uniform Symmetric Array}
\newacronym{UPA}{UPA}{uniform planar array}
\newacronym[]{UE}{UE}{User Equipment}
\newacronym{UL}{UL}{uplink}
\newacronym{V2I}{V2I}{Vehicle-to-Infrastructure}
\newacronym{WP2}{WP2}{Work Package 2}
\newacronym{ETSI}{ETSI}{European Telecommunications Standards Institute}
\newacronym{3GPP}{3GPP}{$3^{\rm rd}$ Generation Partnership Project}
\newacronym{RPixA}{RPixA}{Reconfigurable Pixel Antenna}
\newacronym{RCS}{RCS}{radar cross section}
\newacronym{RV}{RV}{random variable}
\newacronym{TDD}{TDD}{time division duplex}
\newacronym{SVD}{SVD}{singular value decomposition}
\newacronym{SOCP}{SOCP}{Second order cone programming}
\newacronym{SPEB}{SPEB}{squared position error bound}
\newacronym{EFI}{EFI}{effective Fisher information}
\newacronym{EFIM}{EFIM}{effective Fisher information matrix}
\newacronym{BW}{BW}{bandwidth}
\newacronym{GA}{GA}{genetic algorithm}
\newacronym{SDR}{SDR}{semi definite relaxation}
\newacronym{TRP}{TRP}{total radiated power}

\title{Optimizing ISAC MIMO Systems with Reconfigurable Pixel Antennas}
\author{
\IEEEauthorblockN{
Ataher Sams\textsuperscript{1}, 
Yu-Cheng Hsiao\textsuperscript{2}, 
Muhammad Talha\textsuperscript{1}, 
Besma Smida\textsuperscript{1}, and 
Ashutosh Sabharwal\textsuperscript{2}
}
\IEEEauthorblockA{
\textsuperscript{1}\textit{Department of Electrical and Computer Engineering, 
University of Illinois Chicago, Chicago, IL, USA}, \\
\textsuperscript{2}\textit{Department of Electrical
and Computer Engineering, Rice University, Houston, TX, USA} \\
Emails: \{asams3, mtalha7, smida\}@uic.edu, \{yh157, ashu\}@rice.edu
}
}

\maketitle

\begin{abstract}
The integration of sensing and communication demands architectures that can flexibly exploit spatial and electromagnetic (EM) degrees of freedom (DoF). This paper proposes an Integrated Sensing and Communication (ISAC) MIMO framework that uses Reconfigurable Pixel Antenna (RPixA), which introduces additional EM-domain DoF that are electronically controlled through binary antenna coder switch networks. We introduce a beamforming architecture combining this EM and digital precoding to jointly optimize Sensing and Communication. Based on full-wave simulation of pixel antenna, we formulate a non-convex joint optimization problem to maximize sensing rate under user-specific constraints on communication rate. We utilize an Alternating Optimization framework incorporating genetic algorithm for port states of Pixel antennas, and semi-definite relaxation (SDR) for digital beamforming. Numerical results demonstrate that the proposed EM-aware design achieves considerably higher sensing rate compared to conventional arrays and enables considerable antenna reduction for equivalent ISAC performance. These findings highlight the potential of reconfigurable pixel antennas to realize efficient and scalable EM-aware ISAC systems for future 6G networks.

\end{abstract}


\section{Introduction}
The promise of ISAC is to re-use spectrum, hardware and waveforms to sense the environment while maintaining high communication links. To achieve this, we will require more controllable spatial \gls{DoF} than conventional pattern-fixed arrays. An interesting solution is to move beamforming into the EM domain using reconfigurable antennas, which reshape the complex far-field pattern and the effective array manifold. This reconfigurability refers to the ability to alter properties such as frequency, radiation pattern, and polarization. \gls{RPixA}s are a canonical example, using switchable pixels to generate diverse radiation patterns. As highlighted in \cite{antenna_coding, remaa_pixel_antenna_for_MU}, this demands physically-grounded models that capture the actual complex far-field radiation with electronically controllable switching states, rather than relying on idealized, unit-norm abstractions, to reflect real system-level trade-offs.
Beyond \gls{RPixA}, a few families of reconfigurable radiators have emerged that all point to the same modeling idea: a simple ``unit-norm" or ``isotropic" antenna is not physically realizable, and practical "beamforming" should be EM-aware. In parasitic arrays \cite{Reconfig_Parasitic_Heath}, changing passive loads redistributes currents and alters both input impedance and total radiated power, so feasible beamformers are physically constrained and must be designed with the multiport circuit model in mind. In dynamic metasurface antennas (DMAs), wideband behavior (resonance, leakage, frequency selectivity) changes the set of workable precoders and can improve spectral efficiency when co-designed across EM, RF-analog, and baseband \cite{dma_heath}.
In \cite{remaa_pixel_antenna_for_MU}, pixel-enabled electronic movable arrays use pixel coding to emulate aperture displacement without mechanics, delivering multiuser sum-rate gains with fast switching. 
While recent works \cite{tri_hybrid_MU, tri_timescale} formalized tri-hybrid (EM-Analog-Digital) MIMO, they relied on mathematical abstractions and focused solely on communication. These previous works also overlooked \gls{ISAC} scenarios and the co-design of Tx/Rx/EM states with state-dependent power realization. This EM-aware reconfigurability becomes crucial for ISAC systems because, in contrast to fixed-pattern arrays, pixel antennas uniquely provide independent pattern control at each element, which is necessary to simultaneously maximize sensing gain toward radar targets and maintain communication quality to spatially separated users. We address these gaps by providing an \gls{ISAC}-centric, end-to-end pattern-reconfigurable pipeline validated with actual full-wave HFSS simulations.

Our contributions are as follows- (i) Building on multiport network model, we develop an EM-aware channel model that maps binary pixel states to their true, HFSS-simulated complex far-field radiation patterns, rather than relying on idealized array responses. (ii) We pose an \gls{ISAC} optimization that maximizes monostatic sensing SNR while guaranteeing per-user communication rates under transmit power and binary-state constraints at both Tx and Rx. (iii) We propose an \gls{AO} solution pipeline that jointly finds the binary pixel coder states and the digital beamforming. (iv) Finally, through numerical experiments, we show that the EM-aware, pattern-reconfigurable design significantly outperforms conventional arrays and achieves same \gls{ISAC} performance with around 50\% fewer antenna elements.

\section{System Model}

\begin{figure}[th!]
    \centering
    \includegraphics[width=0.9\linewidth]{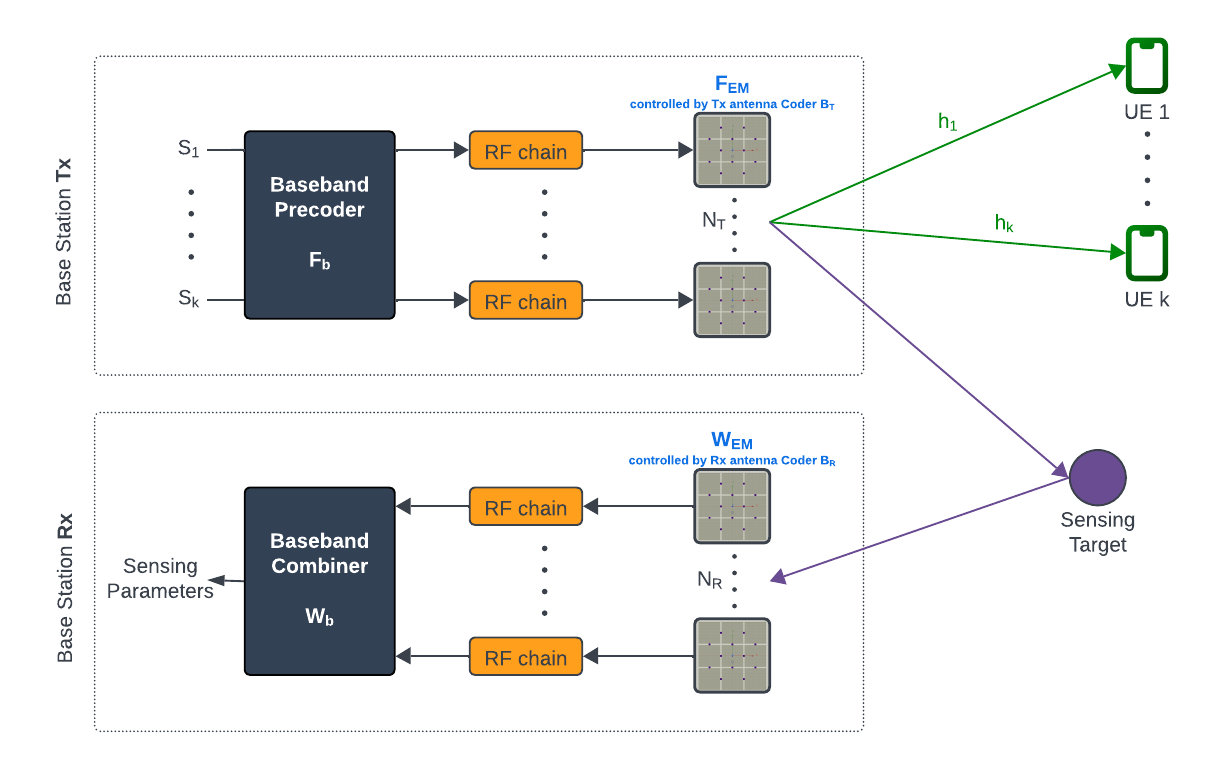}
    \caption{Proposed EM-aware ISAC setting where a base station equipped with a \gls{RPixA} performs joint MU communication and monostatic sensing.}
    \label{fig:channel_model}
\end{figure}

Following the multiport network theory in  \cite{antenna_coding}, we model a RPixA as a ($Q+1$)-port network (see Fig. \ref{fig:pixel_antenna_2point4GHz}), including one active feeding port, denoted `$A$', and $Q$ pixel ports, denoted `$P$'. Its reconfigurability is enabled by a binary antenna coder $\mathbf{b} = [b_1, \ldots, b_Q]^T \in \{0,1\}^{Q}$, where each element $b_q$ controls the switch on pixel port $q$. This coder determines the diagonal load impedance matrix $\mathbf{Z}_{L}(\mathbf{b})  \in \mathbb{C}^{Q\times Q}$, which is applied across $Q$ pixel ports. We adopt the ideal switch model, where the load impedance for the $q$-th port is:
\begin{equation}
        z_{L,q}(b_q) =
    \begin{cases}
        0, & \text{if } b_q=0 \quad \text{(switch ON)} \\
        \infty, & \text{if } b_q=1 \quad \text{(switch OFF)}
    \end{cases}
\end{equation}
The current distribution of RPixA is dependent on the load impedance matrix $\mathbf{Z}_{L}(\mathbf{b})$. Let $i_A$ be the feeding current and $(v_A,\mathbf{v}_P)$ be the antenna and port voltages, respectively. We can relate the currents and voltages by
$$
\begin{bmatrix}\!v_A\\ \mathbf{v}_P\!\end{bmatrix}
= \begin{bmatrix}\!v_A\\ - \mathbf{i}_P \mathbf{Z}_L(\mathbf{b})\!\end{bmatrix} 
= \underbrace{
    \begin{bmatrix}
        z_{AA} & \mathbf{z}_{AP} \\
        \mathbf{z}_{PA} & \mathbf{Z}_{PP}
    \end{bmatrix}}_{\mathbf{Z}}
    \begin{bmatrix}\!i_A\\ \mathbf{i}_P(\mathbf{b})\!\end{bmatrix},
    $$
where the impedance matrix $\mathbf{Z} \in \mathbb{C}^{(Q+1)\times(Q+1)}$ fully characterizes its radiation behavior. Thus, the full-port current vector $\mathbf{i}(\mathbf{b}) \in\mathbb{C}^{(Q+1)}$ is
\begin{equation}
\mathbf{i}(\mathbf{b})=
\begin{bmatrix}
i_A\\[2pt]
\mathbf{i}_P(\mathbf{b})
\end{bmatrix}
=
\begin{bmatrix}
i_A\\[2pt]
-\big(\mathbf{Z}_{PP}+\mathbf{Z}_L(\mathbf{b})\big)^{-1}\mathbf{z}_{PA}\,i_A
\end{bmatrix}
.
\label{eq:multiport-current}
\nonumber
\end{equation}
Similar to \cite{antenna_coding}, we characterize radiation patterns by a far-field vector vector $\mathbf{e_{meas}}(\mathbf{b})$, which is calculated over an $M$-point angular grid using a fixed open-circuit radiation matrix $\mathbf{E}_{\!oc} \in \mathbb{C}^{M \times (Q+1)}$:
\begin{equation}
    \mathbf{e_{meas}}(\mathbf{b}) = \mathbf{E}_{\!oc}\,\mathbf{i}(\mathbf{b}) \in \mathbb{C}^{M\times 1}.
    \label{eq:e_b}
\end{equation}
With this complex far-field vector, which preserves both the magnitude and phase information inherent to patterns generated by a specific coder, $\mathbf{b}$, we can incorporate EM beamforming capabilities into our system model, as shown in Fig. \ref{fig:channel_model}.
\\

\subsection{EM-aware Beamforming Channel Architecture}

In this work, we combine EM beamforming with baseband digital beamforming, leveraging the full DoF to maximize our end goal for ISAC.
\\
\textbf{EM Beamformer $\mathbf{F}_{EM}$:} For the $n$-th antenna, its radiation characteristic is represented by the complex far-field vector $\mathbf{e}_n(\mathbf{b}_n)$, chosen via its coder $\mathbf{b}_n$. The EM beamformer $\mathbf{F}_{EM} \in \mathbb{C}^{MN_T \times N_T}$ is a block diagonal matrix that assembles these individual complex patterns.
\begin{equation}
    \mathbf{F}_{EM}
    = \operatorname{blkdiag}\!\big(\mathbf{e}_1(\mathbf{b}_1),\ldots,\mathbf{e}_{N_T}(\mathbf{b}_{N_T})\big).
\end{equation}
\textbf{Digital Beamforming Architecture:} We consider an ISAC system where a Base Station (BS) with $N_T$ transmit (Tx) and $N_R$ receive (Rx) pixel antennas senses a target and communicates with $K$ users. 
The digital beamformer $\F_b\!\in\!\mathbb{C}^{N_T\times K}$ maps the $K$ data streams $\mathbf{s}$ (with $\mathrm{E}[\mathbf{s}\mathbf{s}^H]\!=\!\mathbf{I}_K$) to the antennas under a total power constraint $\|\F_b\|_F^2\!\le\!P$. 
The transmitted signal is 
\begin{equation}
    \mathbf{x} = \F_b\,\mathbf{s} \in \mathbb{C}^{N_T\times1}.
\end{equation}
To connect the digital and EM beamformers{\footnote{While this model can be readily extended to include analog beamforming, we omit it here to focus on the pattern reconfigurability of \gls{RPixA}. We also assume perfect self-interference mitigation, leaving the inclusion of both aspects as future work.}, we define the overall array response by the geometric steering vector $\mathbf{a}_{\mathrm{g}}(\theta)$. It captures the phase progression due to the array's physical layout, where 
$\mathbf{a}^{(t)}_{\mathrm{g}}(\theta) \in \mathbb{C}^{N_T \times 1}$ and 
$\mathbf{a}^{(r)}_{\mathrm{g}}(\theta) \in \mathbb{C}^{N_R \times 1}$ 
represent the transmit and receive steering vectors, respectively. 
{\small
\begin{equation}
    \mathbf{a}^{(t)}_{\mathrm{g}}(\theta) 
    = \frac{1}{\sqrt{N_T}}
    \big[1, e^{-j\frac{2\pi}{\lambda}d\sin\theta}, \ldots, e^{-j\frac{2\pi}{\lambda}(N_T-1)d\sin\theta}\big]^T.
    \nonumber
\end{equation}
}
With this established, we find the effective complex pattern factor, $c_n(\theta; \mathbf{b}_n)$, for the $n$-th antenna in direction $\theta$ by selecting the corresponding value from its complex pattern vector $\mathbf{e}_{n}(\mathbf{b}_n)$ using a look-angle selection vector $\mathbf{\Pi}(\theta)$ in a similar way presented in \cite{tri_timescale}:
\begin{equation}
    c_n(\theta; \mathbf{b}_n) = \mathbf{\Pi}^T \mathbf{e}_{n}(\mathbf{b}_n) \in \mathbb{C}.
\end{equation}
These complex factors are collected into diagonal matrices, $\mathbf{C}_T(\theta; \mathbf{B}_T)$ and $\mathbf{C}_R(\theta; \mathbf{B}_R)$, respectively. The final, EM-weighted array response is the element-wise product of these complex pattern factors and the geometric steering vectors. From this point forward, $\mathbf{a}_T$ and $\mathbf{a}_R$ will always refer to these complete, EM-weighted vectors.
\begin{align}
    \mathbf{a}_T(\theta;\mathbf{B}_T) &= \mathbf{C}_T(\theta;\mathbf{B}_T)\,\mathbf{a}^{(t)}_{\mathrm{g}}(\theta) \in \mathbb{C}^{N_T \times 1} ,\\
    \mathbf{a}_R(\theta;\mathbf{B}_R) &= \mathbf{C}_R(\theta;\mathbf{B}_R)\,\mathbf{a}^{(r)}_{\mathrm{g}}(\theta) \in \mathbb{C}^{N_R \times 1}.
\end{align}

\subsection{Downlink Communication Channel}
Assume each $k$-th user has single omnidirectional antenna. We focus on a dominant path with an \gls{AoD} $\theta_k$ and complex gain $\alpha_k\in\mathbb{C}$. The array-domain channel row for user $k$ is defined as,
\begin{equation}
    \mathbf{h}_k^H = \alpha_k\, \mathbf{a}_T^H(\theta_k;\mathbf{B}_T) \ \in\ \mathbb{C}^{1\times N_T},
\end{equation}
where $\mathbf{a}_T(\theta_k;\mathbf{B}_T)$ is the complete EM-weighted transmit array response. The full downlink channel matrix for all $K$ users is formed by stacking these rows: $  \mathbf{H}_{\!dl} = \big[\mathbf{h}_1,\, \mathbf{h}_2,\, \ldots,\, \mathbf{h}_K\big]^T \in \mathbb{C}^{K \times N_T}.$
The signal received by the $k$-th user is- 
\begin{equation}
    y_k = \mathbf{h}_k^H \mathbf{x} + n_k = \alpha_k \mathbf{a}_T(\theta_k; \mathbf{B}_T)^H \left( \F_b \mathbf{s} \right) + n_k,
\end{equation}
where $n_k \sim \mathcal{C} \mathcal{N}\left(0, \sigma_k^2\right)$ is the AWGN noise at user $k$. Consequently, the per-user \gls{SINR} is now defined without the analog beamforming matrix:
\begin{equation}
    {\mathrm{SINR}}_k = \frac{\left| \alpha_k \mathbf{a}_T(\theta_k; \mathbf{B}_T)^H \mathbf{f}_{b,k} \right|^2}{\sum_{j \neq k} \left| \alpha_k \mathbf{a}_T(\theta_k; \mathbf{B}_T)^H \mathbf{f}_{b,j} \right|^2 + \sigma_k^2},
\end{equation}
where $\mathbf{f}_{b,k}$ is the $k$-th column of digital beamformer $\F_b$.


\section{Monostatic Sensing Task}
For the sensing task, the objective of the \gls{BS} is to estimate the target direction $\theta_s$ from the target reflections. The round-trip sensing channel $\mathbf{H}_s \in \mathbb{C}^{N_R \times N_T}$, which depends on both coders $\mathbf{B}_T$ and $\mathbf{B}_R$, can be expressed as
\begin{equation}
    \mathbf{H}_s(\theta_s; \mathbf{B}_T, \mathbf{B}_R) = \kappa\beta_s \mathbf{a}_R(\theta_s; \mathbf{B}_R) \mathbf{a}_T(\theta_s; \mathbf{B}_T)^H ,
\end{equation}
where $\beta_s \in \mathbb{C}$ is a complex scalar accounting for path loss and the target's radar cross-section, and $\kappa = \sqrt{N_TN_R}$ is the array scaling factor. 

The transmitted signal $\mathbf{x} \in \mathbb{C}^{N_T \times 1}$ is formed by the digital beamformer $\mathbf{F}_b \in \mathbb{C}^{N_T \times K}$ as $\mathbf{x} = \mathbf{F}_b \mathbf{s}$. The echo signal is received and processed by the digital combiner $\mathbf{W}_{b} \in \mathbb{C}^{N_R \times N_R}$. The received baseband signal at the BS is $\mathbf{y}_s \in \mathbb{C}^{N_R \times 1}$:
\begin{equation}
    \mathbf{y}_s = \mathbf{W}_{b}^H \mathbf{H}_s(\theta_s; \mathbf{B}_T, \mathbf{B}_R) \mathbf{F}_b \mathbf{s} + \tilde{\mathbf{n}},
\end{equation}
where $\tilde{\mathbf{n}} \in \mathbb{C}^{N_R \times 1}$ represents the receiver noise. The quality of the sensing is determined by the Sensing \gls{SNR} of this processed echo, which we denote as:
\begin{equation}
\label{eq:sensing_SNR_digital}
\gamma_{\mathrm{rad}} = \tfrac{|\beta_s|^2\kappa^2}{\sigma_n^2}
\big\| \mathbf{a}_T(\theta_s; \mathbf{B}_T)^H \mathbf{F}_{b} \big\|_F^2 \,
\big\| \mathbf{W}_{b}^H \mathbf{a}_R(\theta_s; \mathbf{B}_R) \big\|_F^2
\end{equation}

\subsection{Comparison with conventional antenna array}
{
    To compare the performance of a pixel antenna array with a conventional antenna array, we look at the norm of the channel matrices of both cases. For the pixel antenna structure, the effective norm of the array response matrix can be given as
    {
\begin{align}
    &\|\mathbf{H}_{s}(\theta_s; \mathbf{B}_T, \mathbf{B}_R)\|_F^2
    = \|\mathbf{a}_R(\theta_s; \mathbf{B}_R)\|^2
       \|\mathbf{a}_T(\theta_s; \mathbf{B}_T)\|^2 \notag\\
    &= \frac{1}{N_T N_R}
       \Bigg(\sum_{n=1}^{N_T}|c_n^T(\theta_s; \mathbf{b}_n^T)|^2\Bigg)
       \Bigg(\sum_{n=1}^{N_R}|c_n^R(\theta_s; \mathbf{b}_n^R)|^2\Bigg).
       \label{eq:norm_pixel_response}
\end{align}
}
For the conventional case, we can write the array response matrix norm as $\norm{\a_R(\theta)\a_T^\herm(\theta)}_\mathrm{F}^2 = \norm{\a_R(\theta)}^2\norm{\a_T(\theta)}^2 = 1$. For pixel antenna we can decouple the transmitter and receiver side to get
    \begin{align}
        \min_{i \in \{T,R\}} \quad & \quad \norm{\a_i(\theta_s;\B_i)}^2 \geq 1\notag\\
        & = \norm{\a_i(\theta_s;\B_i)}^2 \geq 1,\quad \forall i \in \{T,R\} .\label{eq:min_norm_condition}
    \end{align}
    In \eqref{eq:min_norm_condition}, we have effectively decoupled the \gls{TX} and \gls{RX} sides and hence we can look at them individually as  $\norm{\a_i(\theta;\B_i)}^2  = \frac{1}{N_i}\sum_{k = 1}^{N_i}\abs{c_k^i(\theta;\b_k^i)}^2$.
    If we want to have more gain than the conventional antenna array, then the following condition needs to be satisfied
    \begin{align}
        \sum_k\abs{c_k^i (\theta;\b_k^i)}^2 &\geq N_i\label{eq:per_element_cond_seperate_bits},  \qquad i \in \{T,R\}.
    \end{align}
    In other words, the collective gain of all the antennas towards the target direction should be greater than the conventional antenna.
        In order to further expand the \eqref{eq:per_element_cond_seperate_bits}, we consider two strategies to normalize the radiated power of each antenna in order to make a fair comparison against an isotropic conventional antenna: (1) \gls{TRP} normalization and (2) port current normalization, which we will explain shortly. For \gls{TRP} normalization we write \eqref{eq:per_element_cond_seperate_bits} as
        \begin{align}
            \sum_k\abs{c_k^i(\theta,\b_k^i)}^2 
             = \sum_k&\abs{\frac{\Piv^\tran(\theta)\e_{\mathrm{meas}}(\b_k^i)}{\norm{\e_{\mathrm{meas}}(\b_k^i)}_2}}^2 \geq \frac{N_i}{M}, \label{eq:TRP_norm_constraint}
        \end{align}
where, for our 2D azimuthal pattern at fixed elevation, this discrete approximation of TRP is defined as- $\operatorname{TRP}(\mathbf{b})=\frac{1}{M} \sum_{m=1}^M\left|\mathbf{e}_{\text {meas }, m}(\mathbf{b})\right|^2,$
where $\mathbf{e}_{\text{meas}, m}(\mathbf{b})$ is the complex field at the $m$-th sampled angle $\theta_m \in[\theta_{\min}, \theta_{\max}]$ calculated through Equ. \eqref{eq:e_b}. 
This ensures $\frac{1}{M}\sum_{m=1}^{M}|\mathbf{e}_{m}(\mathbf{b})|^2 = 1$ for all coder states $\mathbf{b}$, equivalently $\|\mathbf{e}(\mathbf{b})\|_2^2 = M$. 
For comparison, a conventional isotropic element with uniform angular response satisfies the same unit TRP condition after normalization, establishing a fair performance baseline.

For the current norm, we normalize the $\e_\mathrm{meas}(\b)$ with the norm of the input current to the ports, i.e., $\i(\b)$ as $\e(\b) = \frac{\e_\mathrm{meas}(\b)}{\norm{\i(\b)}}. $  
The rationale behind this is that we normalize the input power to the antenna ports. By using the current norm, the required threshold for better performance for $i \in \{T,R\}$, as compared to a conventional antenna, becomes,
        {\small
        \begin{align}
             \sum_k\abs{c_k^i(\theta,\b_k^i)}^2 & =\sum_k\abs{\frac{\Piv^\tran(\theta)\e_{\mathrm{meas}}(\b_k^i)}{\norm{\i(\b_k^i)}_2}}^2 \geq N_i.
        \end{align}
        }

\section{Optimization problem and solution}
The objective of our work is to design our beamforming matrices as well as pixel antenna switch states such that the sensing performance is maximized subject to given data rate constraints. In particular, from \eqref{eq:sensing_SNR_digital}, that the radar \gls{SNR} given as
\begin{align}
    \hat{\gamma}_{\mathrm{rad}} = \frac{\norm{{\W_b^{\mathrm{H}}}\widehat{\mathbf{H}}_s(\theta_s;\B_T,\B_R)\F_b }_\mathrm{F}^2}{\sigma_b^2} \label{eq:radar_SNR},
\end{align}
where $\widehat{\mathbf{H}}_\mathrm{s} = \a_R(\theta;\B_R)\a_T^\herm(\theta;\B_T)$. With the help of \eqref{eq:radar_SNR}, the optimization problem we aim to solve is
\begin{subequations} \label{P:Original}
    \begin{align}
        & \max_{\W_b,\F_b,\B_R,\B_T} \quad \hat{\gamma}_{\mathrm{rad}} \label{P:Objective_Func}\\
    \textrm{s.t.}&\quad \mathrm{SINR}_k \geq r_k, \quad \forall k \in \setK \label{P:SINR_constraint}\\
   & \norm{\W_b}_\mathrm{F}^2 \leq 1 ,\label{P:W_b_power}\\
   & \norm{\F_b}_\mathrm{F}^2 \leq P,\label{P:F_b_power}\\
   &\mathbf{b}_i^T \in \{0,1\}^Q, \quad \forall i \in \{1,\cdots,N_t \} \label{P:b_T}\\
   &\mathbf{b}_i^R \in \{0,1\}^Q, \quad \forall i \in \{1,\cdots,N_r \}\label{P:b_R}\\
   & \sum_{k = 1}^{N_i}\abs{c_k^i(\theta,\b_k^i)}^2 \geq N_i\quad \forall i \in \{T,R\}. \label{P:per_element_gain_cond}
    \end{align}
\end{subequations}
where \eqref{P:Objective_Func} is radar \gls{SNR} maximization objective function, \eqref{P:SINR_constraint} is the per user \gls{SINR} constraint, \eqref{P:b_T} and \eqref{P:b_R} are the pixel antenna switches state constraints. 

To find the solution of \eqref{P:Original}, we employ the strategy of \gls{AO} and solve this by only assuming one variable active (others constant) at a time. We start with the optimization of $\B_T$ and $\B_R$ and then move towards the digital beamforming optimization\footnote{The reason for choosing this strategy is that it can reduce the number of \gls{AO} iterations required to reach the solution. For instance, if we start with the optimization of $\F_b$ or $\W_b$ for a given random $\H_{\mathrm{s}}$, it might be possible that the given $\B_T$ and $\B_R$ make the vector $\C_T(\theta_s;\B_T)\a_T^\herm(\theta_s)$ orthogonal to optimal $\F_b$ or $\W_b$, which may in return require more number of \gls{AO} iterations.}.


 \subsection{Optimization of $\B_T$ and $\B_R$}
 To solve for both $\B_T$ and $\B_R$ given digital beamformings, we note that both the objective function and users' \gls{SINR} constraints are highly nonlinear in $\B_T$ and $\B_R$ due to constraints \eqref{P:b_T} and \eqref{P:b_R}. Hence, we rely on the \gls{GA} \cite{Genetic_algorithm} to find the optimal $\B_T$ and $\B_R$. We want to emphasize that \gls{GA} has been extensively utilized in pixel antenna design systems \cite{GeneticAlgo_Pixel1, GeneticAlgo_Pixel2}. We use the Lagrangian relaxation to incorporate users' channel strength in the objective function as follows
 \begin{equation}
\max_{\B_T,\B_R} \, \norm{\widehat{\H}_{s}(\theta_s;\B_T,\B_R)}_\mathrm{F}^2 + \sum_k\lambda_k\norm{\a_T(\phi_k;\B_T)}^2
\end{equation}
 Expanding the norm of the sensing channel, we can get $\|\widehat{\mathbf{H}}_s(\theta_s;\B_T,\B_R)\|_F^2
=\|\mathbf{a}_R(\theta;\mathbf{B}_R)\|^2\|\mathbf{a}_T(\theta;\mathbf{B}_T)\|^2$.
 
 As the user channel gain is independent of $\B_R$, we can have two parallel \gls{GA} problems which we solve numerically,
\begin{subequations}\label{G1G2}
\begin{align}
\mathcal{G}_1:\ 
&\max_{\mathbf{B}_T}\ 
\|\mathbf{a}_T(\theta;\mathbf{B}_T)\|^2
+\sum_{k\in\mathcal{K}}\lambda_k \|\mathbf{a}_T(\phi_k;\mathbf{B}_T)\|^2 \\
&\text{s.t.}\ \eqref{P:b_T},\ \eqref{P:per_element_gain_cond}. \notag
\label{P:G1} \\[6pt]
\mathcal{G}_2:\ 
&\max_{\mathbf{B}_R}\ 
\|\mathbf{a}_R(\theta;\mathbf{B}_R)\|^2  \\
&\text{s.t.}\ \eqref{P:b_R},\ \eqref{P:per_element_gain_cond}. \notag
\label{P:G2}
\end{align}
\end{subequations}

\subsection{Optimization of $\F_b$ and $\W_b$}
 To optimize \eqref{P:Original} for a given $\B_T$ and $\B_R$, since $\W_b$ affects only the sensing SNR, it is chosen as the left singular vectors of $\widehat{\H}s(\theta_s)$, i.e., $\overset{*}{\W}_b = \U_{\widehat{\H}_s}$, where $\U_{\widehat{\H}_s}$ are the left singular vectors of $\widehat{\H}_s(\theta_s)$. 
 
 For finding the optimal $\F_b$, we employ the \gls{SDR} method by relaxing the problem by using \gls{SDR} of rank one matrices $\F_{b,k} = \f_{b,k}\f_{b,k}^\herm$. The objective function can then be written as 
      \begin{align}
        \max_{\F_{b,k}\forall k \in \setK} \quad & \quad \tr(\widehat{\H}_s(\theta_s)^\herm\widehat{\H}_s(\theta_s)\sum_k\F_{b,k})         
    \end{align}
    The objective function is now linear in each $\F_{b,k}$. The \gls{SINR} constraints for a user $k$ can be written as
    \begin{align}
        \frac{\abs{\alpha_k}^2}{r_k} \tr(\A_k(\phi_k)\F_{b,k})\notag\\
        \geq \sum_{j\neq k}(\abs{\alpha_{k}}^2\tr(\A_k(\phi_k)\F_{b,j})) + \sigma_k^2, \label{eq:SDR_relaxation_SINR}
    \end{align}
    where $\A_k(\phi_k,\B_T) = \a_T(\phi_k)\a_T^\herm(\phi_k)$. Using this relaxation, we need rank-one solutions, which will introduce nonconvexity in the problem. Relaxing the rank-one constraint, the problem becomes
    \begin{align}\label{P3:SDR_relaxation}
    \max_{\F_{b,k}\,\forall k \in \setK} \quad & \tr\!\left(\widehat{\H}_s^\herm(\theta_s)\widehat{\H}_s(\theta_s)\sum_k \F_{b,k}\right) \\
    \text{s.t.} \quad 
    & \tr\!\left(\sum_k \F_{b,k}\right) \leq P, \quad \text{and} \quad \eqref{eq:SDR_relaxation_SINR}. \notag 
\end{align}

    The relaxed problem \eqref{P3:SDR_relaxation} may not be tight; the optimal solution of \eqref{P3:SDR_relaxation} can give us solutions with more than rank-one. However, following the strategy defined in \cite{SDR_relaxation}, we can get the rank-one optimal solutions, from which we can construct the optimal $\overset{*}{\F}_{b}$.  We summarize the overall optimization framework in Algorithm \ref{alg:cap}.
\begin{algorithm}[t!]
\caption{Pixel and beamforming optimization}\label{alg:cap}
\begin{algorithmic}[1]
\REQUIRE $\theta,\phi_1,\phi_2,\cdots,\phi_K,I $
\ENSURE $\B_T,\B_R,\W_b,\F_b$
\STATE $i \gets 1$
\STATE $\B_T$ and $\B_R \gets$ solve \eqref{G1G2} 
\STATE $\W_b \gets \frac{\U_{\H_\rm{rad}}}{\norm{\U_{\H_\rm{rad}}}_\rm{F}}$
\STATE $\{\F_{b,k} \forall k \in \setK\}\gets$ solve \eqref{P3:SDR_relaxation} 
\FOR{$k \in \setK$}
\IF{$\mathrm{rank}(\F_{b,k}) \neq 1$}
\STATE $\f_{b,k} \gets \left(\h_k^\herm \F_{b,k}\h_k\right)^{-1/2}\F_{b,k}\h_k$
\ELSE
\STATE $\F_{b,k} = \lambda_k\q_k\q_k^\herm$ and  $\f_{b,k} \gets \lambda_k\q_k$
\ENDIF
\ENDFOR
\end{algorithmic}
\end{algorithm}

\section{Numerical Results}

\begin{figure}
    \centering
    \includegraphics[width=0.9\linewidth]{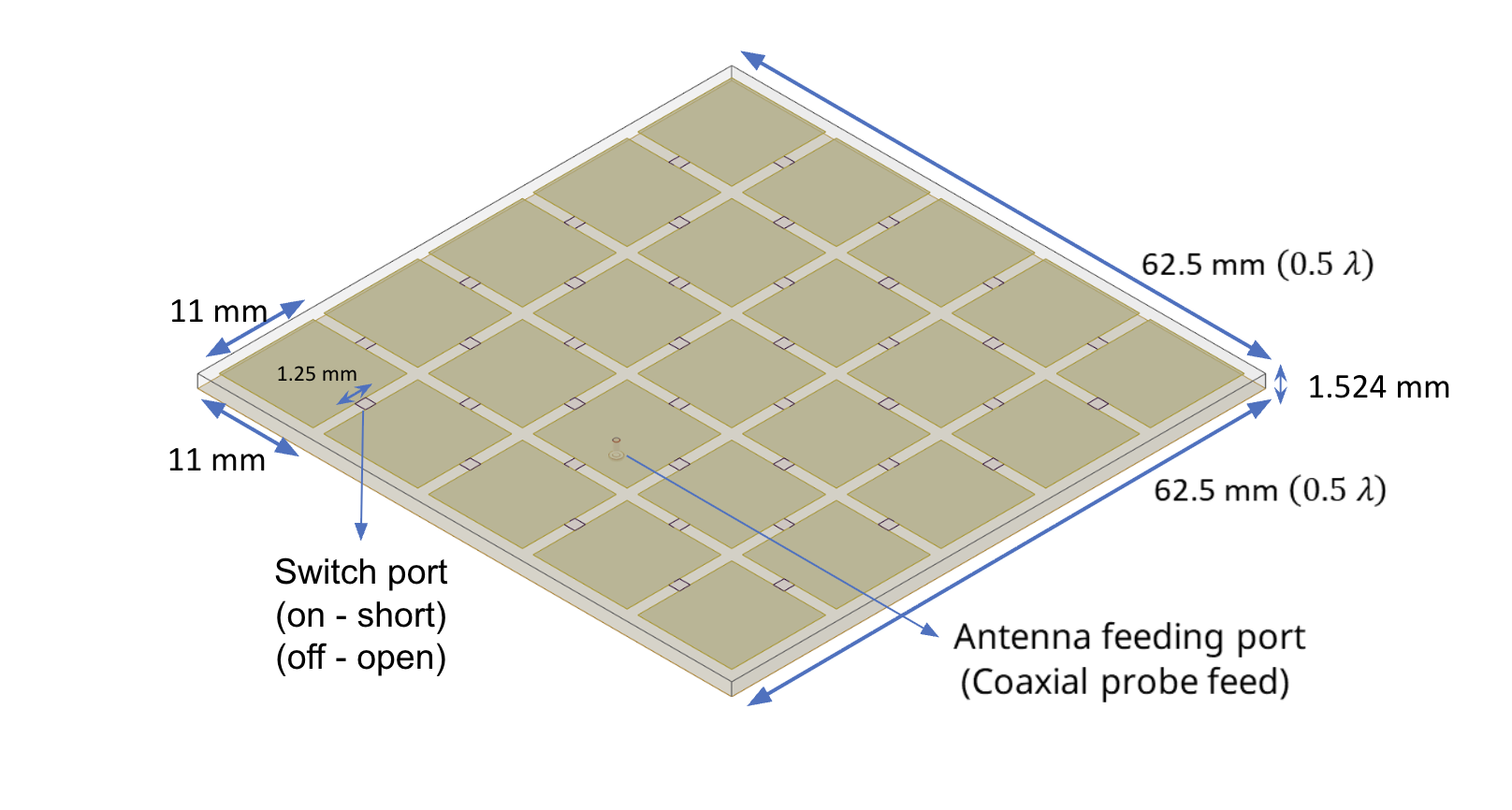}
    \caption{Geometry of proposed pixel antenna featuring one coaxial feeding port and $Q=40$ pixel ports}
    \label{fig:pixel_antenna_2point4GHz}
\end{figure}
\textbf{Reconfigurable Pixel Antenna Design.} We present a $5\times 5$ pixel antenna for 2.4 GHz in Fig.  \ref{fig:pixel_antenna_2point4GHz} with pattern reconfigurability via electronically controlled switches. The dimension of aperture is $0.5\lambda \times 0.5\lambda$ ($\lambda = 125$ mm), includes metallic pixels (11 mm × 11 mm) separated by 1.25 mm gaps. The Rogers RO4003 is selected for the substrate due to its low dielectric loss. There are $40$ RF switches ($2^{40}$ possible states) between pixels encoded by a binary coder $\mathbf{b} \in \{0,1\}^{40}$. Excitation is provided by a coaxial probe optimized for 50 $\Omega$ impedance matching.
Fig. \ref{fig:radiation_pattern} illustrates pattern reconfigurability of a \gls{RPixA} for five random binary coder states under the two normalization approaches. This highlights the benefit of goal-oriented reconfigurable pattern selection and ensures a fair comparison with isotropic antennas.

 \begin{figure}[t]
  \centering
  \setlength{\tabcolsep}{2pt}
  \begin{tabular}{@{}c@{}}
    \includegraphics[width=0.9\linewidth]{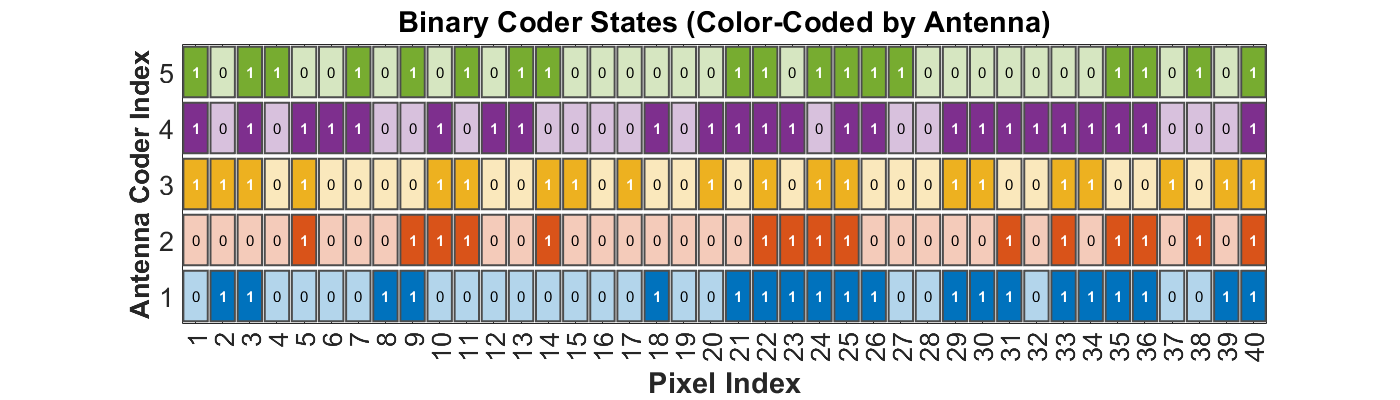} \\[-1.5mm]
    \begin{tabular}{@{}cc@{}}
      \includegraphics[width=0.45\linewidth]{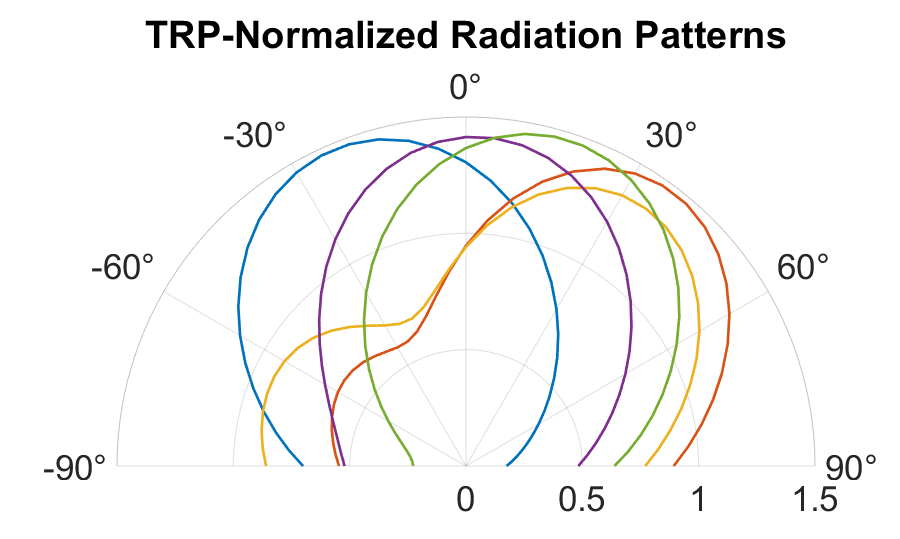} &
      \includegraphics[width=0.45\linewidth]{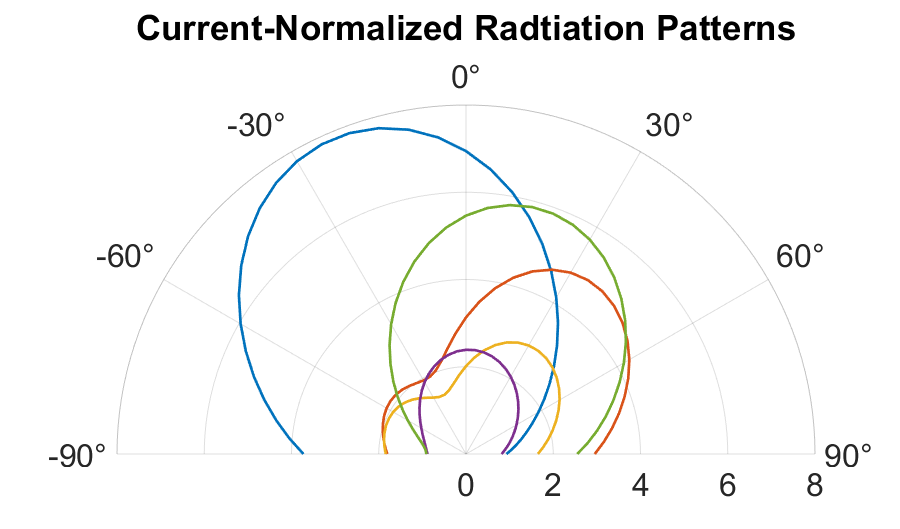}
    \end{tabular}
  \end{tabular}
  \caption{ Radiation patterns for five random Antenna coder of one single \gls{RPixA}; (a) Binary coder states, where light color denotes $0\rightarrow\text{switch on}$, and dark color denotes $1\rightarrow\text{switch off}$ (b) TRP-normalized patterns; (c) current-normalized patterns. Colors are consistent across all panels.}
  \label{fig:radiation_pattern}
\end{figure}


\textbf{Simulation Setup and Parameters.} 
The simulation framework is implemented in MATLAB, utilizing full-wave electromagnetic data from HFSS 2025 R1. Key system parameters include a 2.4 GHz carrier frequency ($f_c$) with $\lambda/2$ element spacing (62.5 mm), a total transmit power ($P_{\text{total}}$) of 20 dBW, and a uniform noise power of -80 dBm for both users and sensing. The target distance is 40 m, with $\text{PathLoss}(d) = \text{FSPL} + 10n\log_{10}(d) + \sigma$ (where $n=2$ and $\sigma=2.9$ dB). 
For the Genetic Algorithm, a population size of 300 and a maximum of 200 generations are used.

\begin{figure}[t]
    \centering
    \setlength{\tabcolsep}{1pt}
    \begin{tabular}{@{}cc@{}}
        \includegraphics[width=0.48\linewidth]{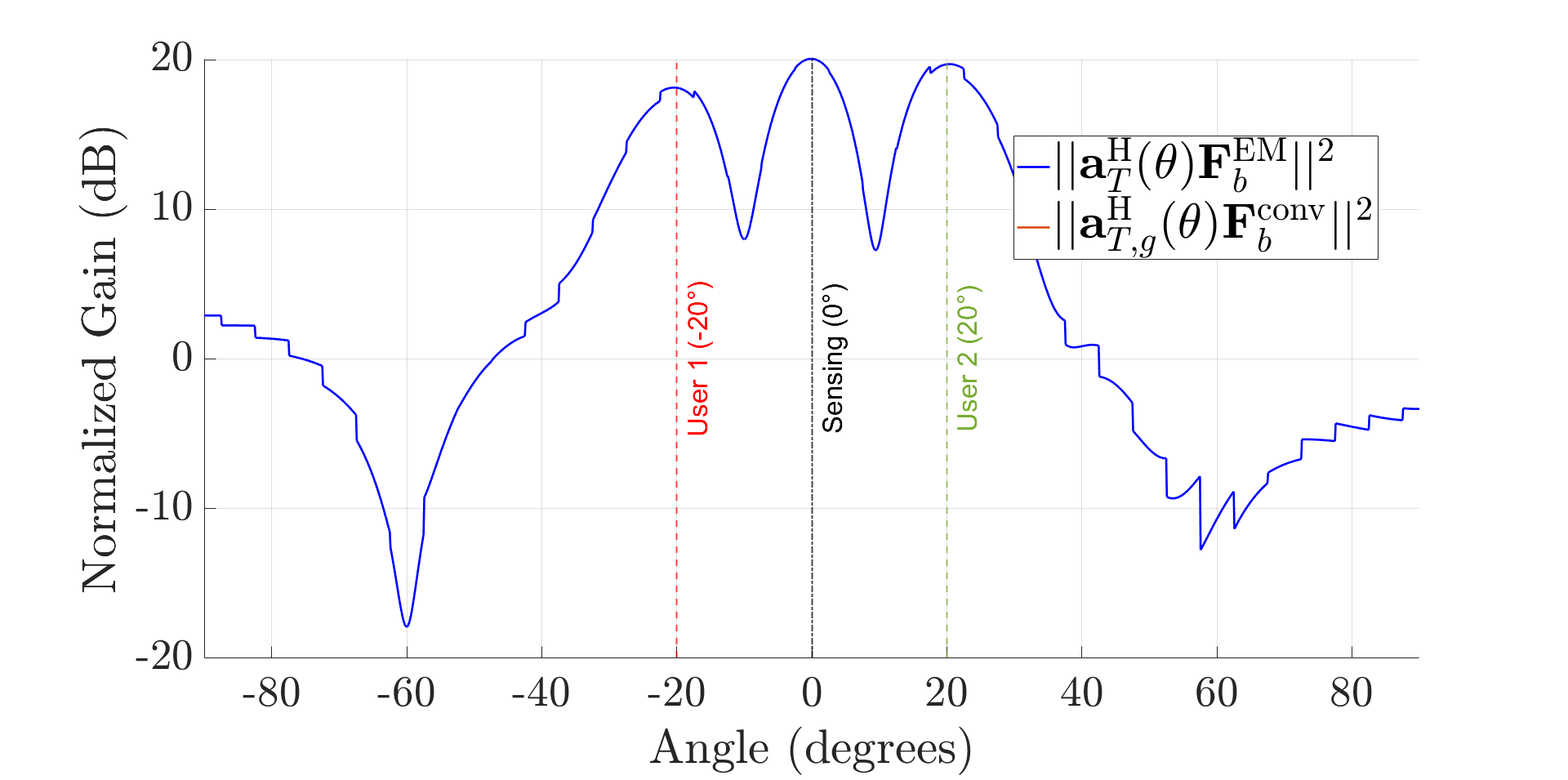} &
        \includegraphics[width=0.48\linewidth]{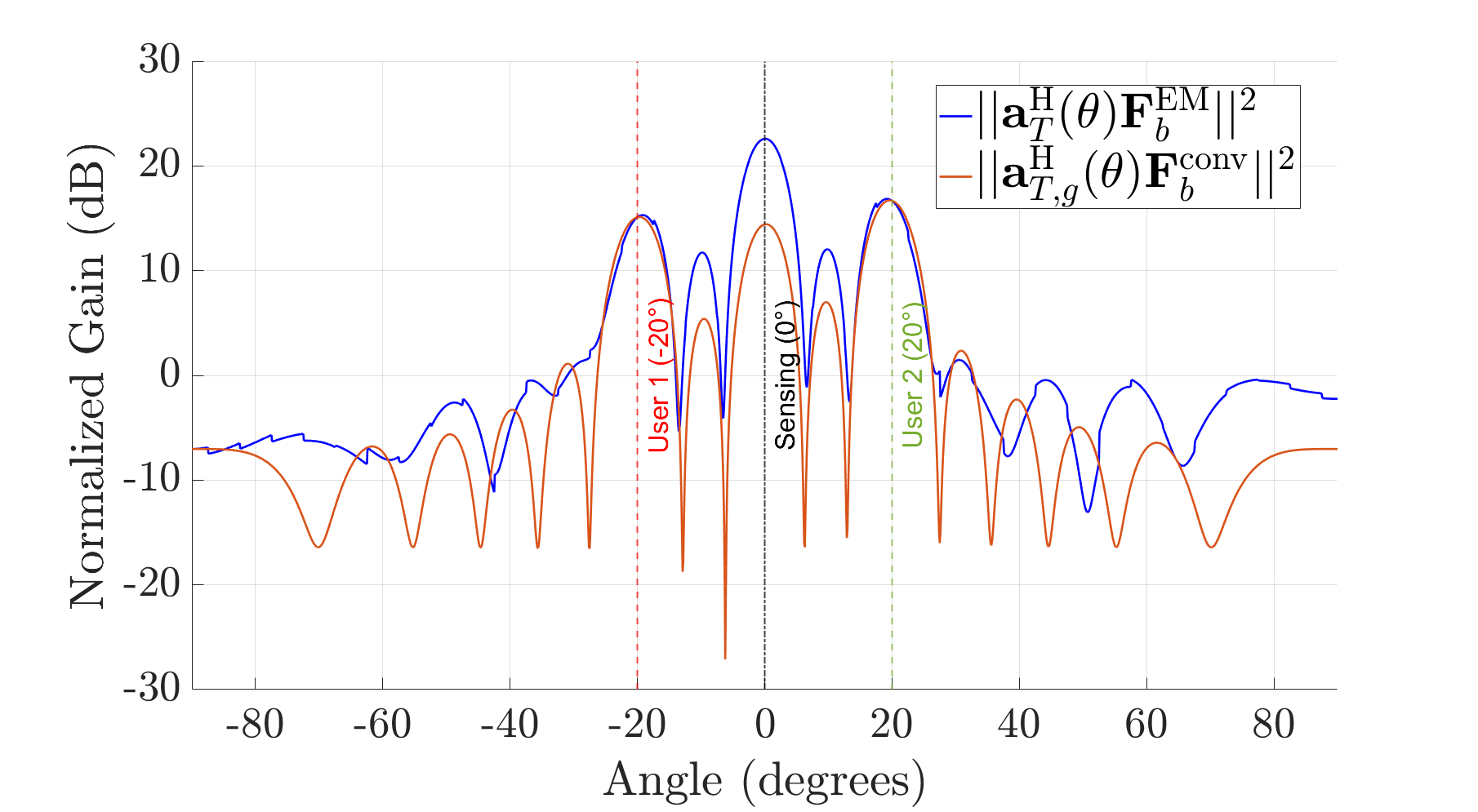} \\
        (a) $N_T = N_R = 8$ & (b) $N_T = N_R = 16$ \\
        \multicolumn{2}{c}{
        \includegraphics[width=0.5\linewidth]{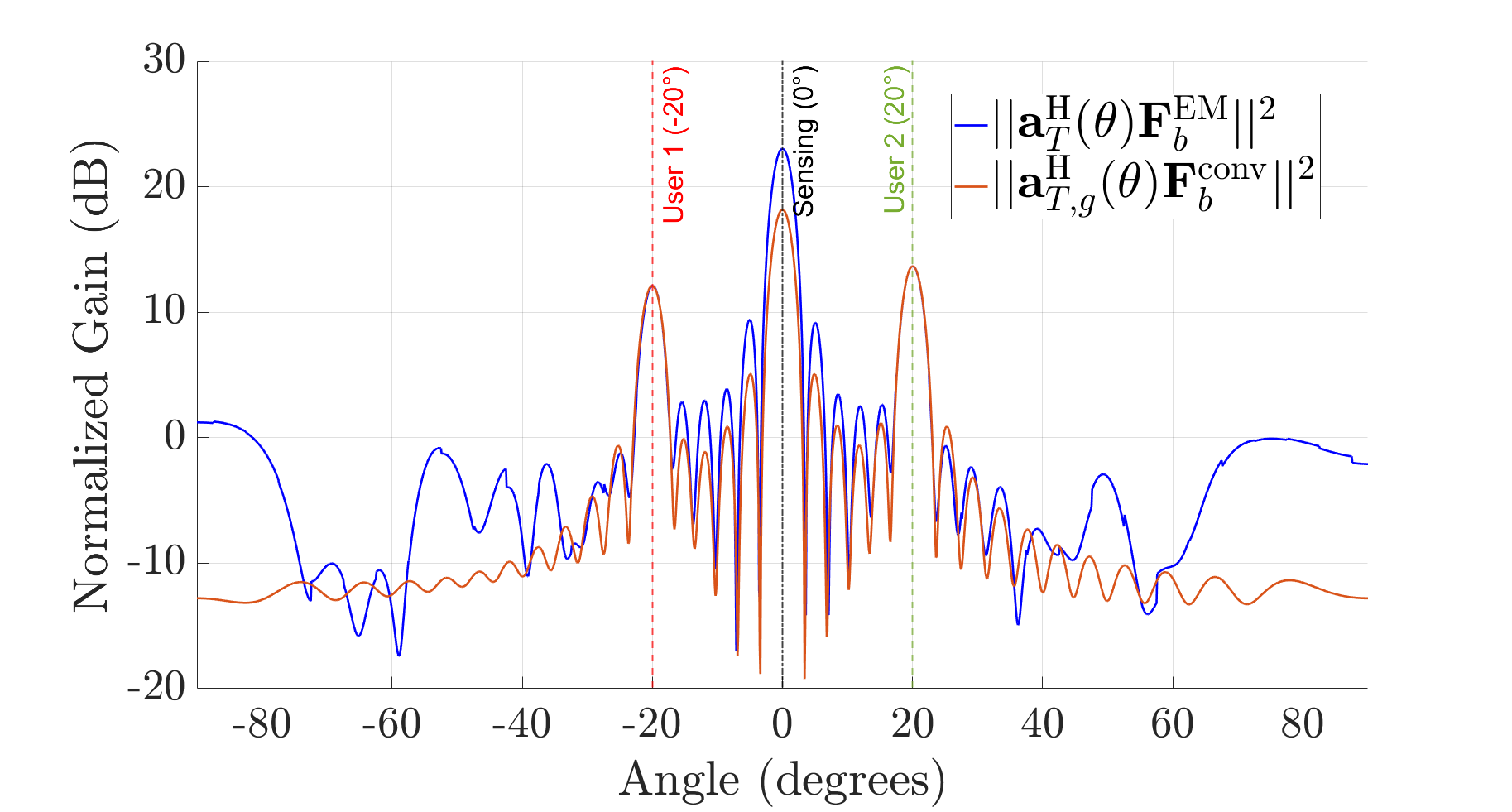}}\\
        \multicolumn{2}{c}{(c) $N_T = N_R = 32$}
    \end{tabular}
    \vspace{-2mm}
    \caption{Optimized Tx beampatterns for different array sizes. Blue: EM-optimized ($|\mathbf{a}_T^{\mathrm{H}}(\theta)\mathbf{F}b^{\mathrm{EM}}|^2$); red: conventional ($|\mathbf{a}_{T,\mathrm{g}}^{\mathrm{H}}(\theta)\mathbf{F}_b^{\mathrm{conv}}|^2$). Dashed lines denote User 1 ($-20^\circ$), sensing target ($0^\circ$), and User 2 ($20^\circ$).}
    \vspace{-2mm}
    \label{fig:beampatterns_combined}
\end{figure}

\textbf{Beampattern Analysis.} Fig.~\ref{fig:beampatterns_combined} compares the optimized Tx beampatterns for 8, 16, and 32-element arrays using our EM-aware \gls{RPixA} model against a conventional isotropic model. With a sensing target at $0^{\circ}$, users at $[-20^{\circ}, 20^{\circ}]$, the EM-optimized design consistently achieves higher peak gain with three distinct main lobes. The conventional design fails to find a feasible solution for $N=N_T=N_R=8$.  The EM-aware design shows a 29\% higher sensing rate for $N=16$ (17.85 vs. 13.8 bps/Hz) and a 17\% improvement for $N=32$ (19.93 vs. 17.06 bps/Hz), all while satisfying communication rate of 10 bps/Hz.
\begin{figure}[htbp]
    \centering
    \begin{subfigure}[b]{0.46\linewidth}
        \centering
        \includegraphics[width=\linewidth]{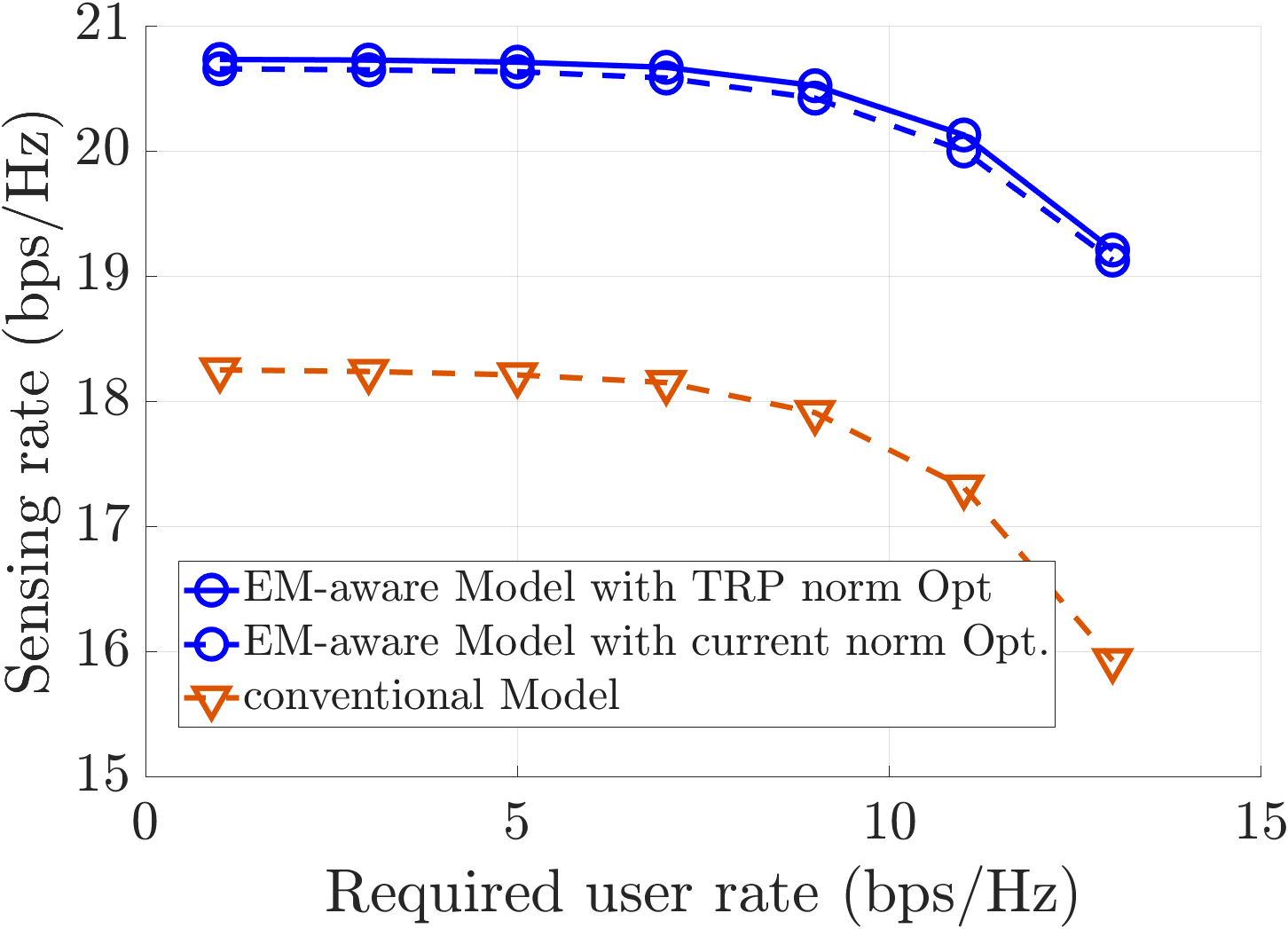}
        \caption{}
        \label{fig:S_rate_vs_reqSINR}
    \end{subfigure}
    \hfill
    \begin{subfigure}[b]{0.48\linewidth}
        \centering
        \includegraphics[width=\linewidth]{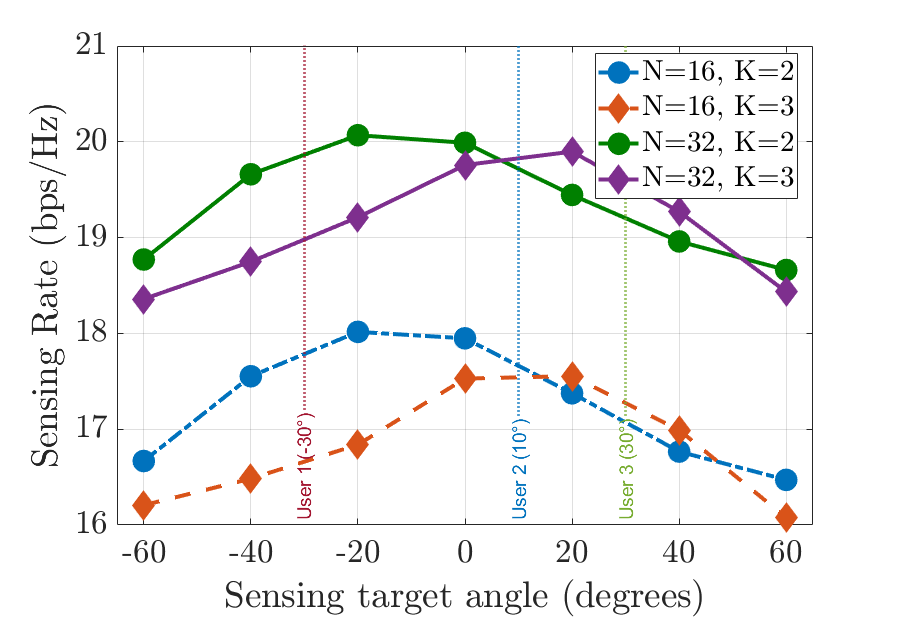}
        \caption{}
        \label{fig:S_rate_vs_sensingAngle}
    \end{subfigure}
    \caption{Performance comparison of sensing rates with (a) varying user SINR and (b) varying target angle.}
    \label{fig:S_rate_combined}
\end{figure}

\textbf{Sensing Rate and Efficiency Analysis.} Fig.~\ref{fig:S_rate_vs_reqSINR} evaluates sensing rate versus user rate requirements (1–14 bps/Hz) for a 32-element array with users at $[-20^\circ, 20^\circ]$ and a target at $0^\circ$. Both EM-aware normalization methods (TRP and port current) are nearly identical and consistently outperform the conventional model; so only the \gls{TRP}-normalized results are used in the subsequent plots. The sensing rate is stable for low user rates ($R_k < 8$ bps/Hz) but degrades at higher constraints, as the optimization must allocate more power to meet the exponential communication SINR requirement ($r_k = 2^{R_k} - 1$), confirming the fundamental \gls{ISAC} trade-off. Subsequently, Fig.~\ref{fig:S_rate_vs_sensingAngle} illustrates sensing rate versus target angle for $N=16$ and $N=32$ arrays. For two users ($K=2$, at $[-30^\circ, 10^\circ]$), peak rates occur near $-20^\circ$ (20.1 bps/Hz for $N=32$, 18.0 bps/Hz for $N=16$). For three users ($K=3$, at $[-30^\circ, 10^\circ, 30^\circ]$), peaks are near $20^\circ$ (19.8 bps/Hz for $N=32$, 17.6 bps/Hz for $N=16$). This confirms that maintaining high sensing rates is more difficult when the sensing targets are far from the users.
\begin{figure}
    \centering
    \includegraphics[width=0.7\linewidth]{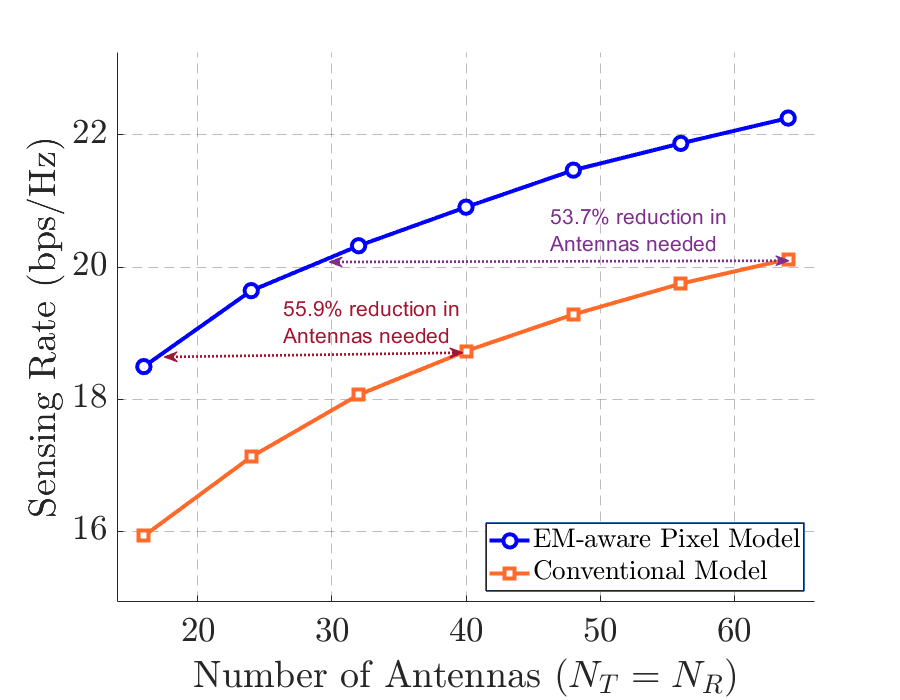}
    \caption{Sensing rate vs array size comparison between EM-aware pixel model and conventional model for user rate constraint of 8 bps/Hz.}
    \label{fig:efficiency}
\end{figure}

Leveraging EM-aware beamforming reduces the required RF chains and antennas for equivalent performance. Figure~\ref{fig:efficiency} demonstrates this by plotting sensing rate versus array size for a two-user scenario (8 bps/Hz rate constraint). The EM-aware pixel model consistently outperforms the conventional design, achieving an average sensing rate improvement of 2.28 bps/Hz. Significantly, the EM-optimized approach enables substantial hardware reduction. For example, only 30 pixel antennas are required to match the 20.12 bps/Hz sensing rate of a conventional 64-element array, representing a 53.7\% reduction in number of antenna and RF-chains. This antenna reduction directly lowers cost and power consumption while maintaining identical \gls{ISAC} performance, demonstrating the scalability and robustness of the reconfigurable pixel antenna approach.

\section{Conclusion}
We presented an EM-aware optimization framework for \gls{ISAC} utilizing \gls{RPixA}. By bridging full-wave EM simulation with digital precoder design, we unlock new \gls{DoF} that were previously unattainable in a conventional fixed pattern antenna array. Through binary antenna coders and a joint optimization pipeline, the work demonstrates how EM-level reconfigurability enhances \gls{ISAC} performance. Future work will focus on developing faster and more efficient optimization algorithms, reducing the pixel-state search space, and incorporating reflection coefficient and coupling effects for more accurate EM modeling. Additionally, extending the framework to multi-band and polarization-reconfigurable designs will introduce a new frontier for 6G.

\bibliographystyle{ieeetr}
\bibliography{reference}

\end{document}